\documentclass[fleqn,usenatbib]{mnras}

\usepackage{newtxtext,newtxmath}

\usepackage[T1]{fontenc}

\DeclareRobustCommand{\VAN}[3]{#2}
\let\VANthebibliography\thebibliography
\def\thebibliography{\DeclareRobustCommand{\VAN}[3]{##3}\VANthebibliography}

\usepackage{graphicx}	
\usepackage{amsmath}	
\usepackage{bm}
\usepackage{orcidlink}

\newcommand{\dd}{\mathrm{d}}

\newcommand{\Var}{\operatorname{Var}}
\newcommand{\Cov}{\operatorname{Cov}}
\newcommand{\calI}{\mathcal{I}}
\newcommand{\calR}{\mathcal{R}}
\newcommand{\boldy}{\boldsymbol{y}}
\newcommand{\boldt}{\boldsymbol{t}}

\usepackage{xcolor}
\usepackage[normalem]{ulem} 

\definecolor{revadd}{RGB}{0,140,90}
\definecolor{revdel}{RGB}{200,40,40}
\definecolor{revcom}{RGB}{140,0,180}

\newif\ifrev
\revtrue          

\ifrev
  
  \newcommand{\revdel}[1]{\textcolor{revdel}{\sout{#1}}}
  
  \newcommand{\revcom}[1]{%
    \textcolor{revcom}{\textbf{[}#1\textbf{]}}}
\else
  
  \newcommand{\revdel}[1]{}
  
  \newcommand{\revcom}[1]{}
\fi

\title[Optimal Magnification]{Optimal galaxy magnification: generalized fundamental-plane residuals and number counts are all you need}

\author[Kun Xu et al.]{
Kun Xu\orcidlink{0000-0002-7697-3306},$^{1}$\thanks{E-mail: kunxu@sas.upenn.edu}
and Ravi K. Sheth\orcidlink{0000-0002-2330-0917}$^{1}$
\\
$^{1}$ Center for Particle Cosmology, Department of Physics and Astronomy, University of Pennsylvania, Philadelphia, PA 19104, USA
}

\date{Accepted XXX. Received YYY; in original form ZZZ}

\pubyear{\the\year{}}

\begin{document}
\label{firstpage}
\pagerange{\pageref{firstpage}--\pageref{lastpage}}
\maketitle

\begin{abstract}
Weak-lensing magnification complements cosmic shear through its effects on galaxy fluxes, sizes, and number counts. We derive the optimal magnification statistics from galaxy properties and show that the optimal measurement can be decomposed into an optimal mean statistic and total number counts. The optimal mean statistic can be constructed from the residual distribution of a generalized fundamental plane (GFP), $F(i,\bm{X})$, which predicts galaxy size or, equivalently, flux from surface brightness $i$ and additional lensing-invariant properties $\bm{X}$. The function $F$ can take any form that reduces the intrinsic scatter and thereby improves magnification precision. For a Gaussian residual distribution, the optimal mean statistic reduces to the inverse-variance-weighted residual, or simply the residual for constant scatter. Together with total number counts, this statistic captures all the magnification Fisher information under Poisson statistics, achieving the same precision as a full-likelihood analysis. This provides a simple and physically intuitive optimal compression of the full magnification likelihood, avoiding the need to model the potentially high-dimensional galaxy-property distribution, particularly when additional observables are incorporated to reduce the GFP scatter. Our results demonstrate that GFP residuals and number counts provide an optimal combination for weak-lensing magnification measurements in galaxy surveys.
\end{abstract}




\section{Introduction}

Weak gravitational lensing provides a powerful probe of the matter distribution in the Universe \citep{2001PhR...340..291B}. It produces two observable effects on background galaxies: shear, which distorts galaxy shapes, and magnification, which changes their apparent sizes, fluxes, and number densities. While weak-lensing measurements have traditionally relied on shear, magnification provides complementary information and is subject to different observational and astrophysical systematics.

Weak-lensing magnification can be measured through several observables. The most widely used approach is magnification bias, which measures changes in galaxy number counts caused by the competing effects of flux amplification and solid-angle dilation \citep{2005ApJ...633..589S,2009A&A...507..683H}. Since the magnification response depends on galaxy luminosity, optimally weighting galaxies according to their number-count slopes can improve the signal-to-noise ratio \citep{2002A&A...386..784M,2005PhRvL..95x1302Z,2016MNRAS.456.2518G}. Alternatively, magnification can be measured directly from changes in galaxy sizes and fluxes \citep{2010MNRAS.405.1025M,2013MNRAS.433L...6H,2024ApJ...973..102X}. In particular, the fundamental plane (FP) of early-type galaxies provides a relation between galaxy size and lensing-invariant properties, allowing the intrinsic size to be predicted with reduced scatter \citep{1995ApJ...451...60B,2006ApJ...648L..17B,2014ApJ...780L..16H,2026arXiv260909151X}. The FP residuals have also been combined with galaxy number counts to improve magnification measurements \citep{2020MNRAS.496.2998F}. More generally, the joint distribution of galaxy sizes and magnitudes has been used to construct a Bayesian likelihood for magnification inference \citep{2012ApJ...744L..22S,2015MNRAS.452.1202A}.

Despite these developments, it remains unclear which observables are sufficient to extract all the available magnification information. Optimizing individual measurements does not necessarily guarantee that their combination is optimal, while a full-likelihood analysis requires modelling the potentially high-dimensional distribution of galaxy properties. Incorporating additional lensing-invariant properties can improve magnification precision by better characterizing the intrinsic distribution of galaxy sizes and fluxes, but also increases the dimensionality and complexity of the likelihood. It is therefore important to understand whether the full information can be compressed into a small number of simple, physically motivated statistics.

In this Letter, we derive the optimal magnification estimators for general galaxy-property distributions. We show that the optimal summary statistic can be decomposed into an optimal mean statistic and total galaxy number counts, and that it achieves the same Fisher information as the full likelihood under Poisson statistics. We further demonstrate that the optimal mean statistic can be constructed from the conditional generalized fundamental-plane (GFP) residual distribution. For a Gaussian GFP residual distribution, it reduces to the inverse-variance-weighted GFP residual, or simply the GFP residual for constant scatter. Thus, the GFP residual and total number counts are sufficient to capture all the magnification information, providing a simple and physically intuitive alternative to high-dimensional likelihood modelling.

\section{Weak lensing general}\label{sec:WL_gen}
Gravitational lensing maps the source-plane position $\boldsymbol{\theta}^{S}$ to the image-plane position $\boldsymbol{\theta}^{I}$. The local lensing distortion is described by the Jacobian matrix
\begin{equation}
    \mathbf{A}
    =
    \frac{\partial \boldsymbol{\theta}^{S}}{\partial \boldsymbol{\theta}^{I}}
    =
    \begin{pmatrix}
        1-\kappa-\gamma_1 & -\gamma_2 \\
        -\gamma_2 & 1-\kappa+\gamma_1
    \end{pmatrix}\,,
\end{equation}
where $\kappa$ is the convergence and $\gamma_1$ and $\gamma_2$ are the two components of the shear, with $\gamma=\gamma_1+i\gamma_2$. The magnification is given by the inverse absolute determinant of the Jacobian matrix,
\begin{equation}
\mu
=
\left|\det\frac{\partial \boldsymbol{\theta}^{I}}{\partial \boldsymbol{\theta}^{S}}\right|
=
\frac{1}{|(1-\kappa)^2-|\gamma|^2|}\,,
\label{eq:mu}
\end{equation}
which describes the change in the observed solid angle. In the weak-lensing limit, where $|\kappa|\ll1$ and $|\gamma|\ll1$, the magnification reduces to $\mu\simeq1+2\kappa$. According to Liouville's theorem, gravitational lensing conserves surface brightness~\citep{2001PhR...340..291B}.

For galaxies, the two observables directly affected by magnification are flux and size. The observed flux is commonly expressed in terms of apparent magnitude, which transforms under weak lensing as
\begin{equation}
    m \rightarrow m-\frac{5}{\ln 10}\kappa\,.
\end{equation}
For size, we consider the circularized effective radius $R_{\rm e}\propto\sqrt{\mathcal{A}_{\rm e}}$, where $\mathcal{A}_{\rm e}$ is the area enclosed by the half-light ellipse. Since the area is unaffected by shear at leading order, the logarithmic size $r\equiv\log_{10}R_{\rm e}$ transforms as
\begin{equation}
    r\rightarrow r+\frac{\kappa}{\ln 10}\,.
\end{equation}
These transformations also demonstrate the conservation of surface brightness. In particular, the logarithmic effective surface brightness,
\begin{equation}
    i\equiv\log_{10}I_{\rm e}
    =c_m-0.4m-2r\,,\label{eq:SB}
\end{equation}
is invariant under magnification, where $c_m$ is a constant that absorbs the choice of units. Although the observed surface brightness is also affected by cosmological (Tolman) dimming and $k$-corrections, these effects are independent of $\kappa$, and therefore do not affect the magnification response considered here.

Moreover, galaxy samples are typically selected based on observed properties, so magnification can move galaxies into or out of the sample. Consider an unlensed galaxy population with differential number density $n_0(\bm{y})$, where $\bm{y}=(r,m,\bm{X})$ and $\bm{X}$ denotes a vector of lensing-invariant properties other than surface brightness $i$, such as velocity dispersion and colour. We define the magnification response vector as
\begin{equation}
    \bm{t}\equiv\frac{\partial\bm{y}}{\partial\kappa}
    =a(1,-5,\bm{0})\,,\qquad a\equiv\frac{1}{\ln 10}.
\end{equation}
For a selection function $S(\bm{y})$, the observed differential number density under magnification is, to first order in $\kappa$,
\begin{equation}
    n_\kappa^S(\bm{y})
    \simeq(1-2\kappa)S(\bm{y})
    n_0(\bm{y}-\kappa\bm{t})\,.
    \label{eq:population}
\end{equation}
The factor $(1-2\kappa)$ accounts for the dilution of the observed number density due to the expansion of the solid angle by a factor $\mu\simeq1+2\kappa$, while the shift in $n_0$ describes the change in observed galaxy properties and the resulting selection effects.

In practice, galaxy properties may be measured in multiple photometric bands, and the selection may depend on measurements across these bands. The formalism can be straightforwardly generalized by replacing $(r,m)$ with the corresponding vectors $(\bm{r},\bm{m})$. For simplicity, we restrict the following discussion to a single band.

\section{Mean Statistics}\label{sec:mean}
In this section, we derive the optimal magnification estimator based on the sample mean of a general galaxy observable. We show that, when the GFP residuals follow a Gaussian distribution, the inverse-variance-weighted residual is the unique optimal observable, up to an affine transformation. We further demonstrate that this estimator achieves the same statistical precision as a full-likelihood analysis.

\subsection{General optimal estimator}
Consider a galaxy observable $f(\boldy)$. The expectation value of its sample mean is
\begin{equation}
    \langle f\rangle_\kappa
    =\frac{\int\dd\boldy\,f(\boldy)\,n_\kappa^S(\boldy)}
           {\int\dd\boldy\,n_\kappa^S(\boldy)}\,.
    \label{eq:mean_expectation}
\end{equation}
We define the magnification response of the sample mean as
\begin{equation}
    \calR_f
    \equiv\left.\frac{\partial\langle f\rangle_\kappa}
                         {\partial\kappa}\right|_{0}\,.
    \label{eq:Rf_define}
\end{equation}
Here, $f$ is held fixed as a function of the observed galaxy properties when differentiating with respect to $\kappa$, so the lensing dependence enters entirely through $n_\kappa$. 
To evaluate this response, we introduce the density score along the lensing direction,
\begin{equation}
    s(\boldy)
    \equiv
    \left.\frac{\partial}{\partial\kappa}
    \ln n_0(\boldy-\kappa\boldt)\right|_{0}
    =-\boldt\cdot\nabla\ln n_0(\boldy)\,.
\end{equation}
From equation~(\ref{eq:population}), we then have
\begin{equation}
    \left.\frac{\partial n_\kappa^S}{\partial\kappa}
    \right|_{0}
    = \big(s(\boldy)-2\big)\,S(\boldy)\, n_0(\boldy)\,.
\end{equation}
Substituting this into equation~(\ref{eq:Rf_define}) gives
\begin{align}
    \calR_f
    &=\langle f(s-2)\rangle_S
    -\langle f\rangle_S\langle s-2\rangle_S \nonumber\\
    &=\Cov_S(f,s)\,,
    \label{eq:response_covariance}
\end{align}
where $\langle\cdot\rangle_S$ and $\Cov_S$ denote the mean and covariance over the unlensed selected population, respectively. 

For a sample of $N$ galaxies, neglecting measurement errors in the galaxy properties $\boldy$, the inverse variance of the magnification estimator based on the sample mean of $f$ is
\begin{equation}
    \calI_f^{\rm mean}
    =N\frac{\calR_f^2}{\Var_S(f)}
    =N\frac{\Cov_S(f,s)^2}{\Var_S(f)}\,.
    \label{eq:fisher_f}
\end{equation}
By the Cauchy--Schwarz inequality,
\begin{equation}
    \Cov_S(f,s)^2
    \leq\Var_S(f)\Var_S(s),
\end{equation}
which gives an upper bound on the information,
\begin{equation}
    \calI_f^{\rm mean}\leq N\Var_S(s)\,.
    \label{eq:information_bound}
\end{equation}
Equality holds if and only if
\begin{equation}
    f(\boldy)=A\,s(\boldy)+B,\qquad A\neq0,
    \label{eq:uniqueness}
\end{equation}
within the selected population. Therefore, the unique optimal observable, up to an affine transformation, is
$f_{\rm opt}^{\rm mean}(\boldy)\propto s(\boldy)$.

Notably, the normalized single-galaxy likelihood and its score are
\begin{align}
    \mathcal{L}^{\rm mean}(\kappa|\boldy)
    &=\frac{n_\kappa^S(\boldy)}
    {\int\dd\boldy'\,n_\kappa^S(\boldy')}
    =\frac{S(\boldy)n_0(\boldy-\kappa\boldt)}
    {\int\dd\boldy'\,S(\boldy')n_0(\boldy'-\kappa\boldt)}, \nonumber\\
    \left.\frac{\partial\ln\mathcal{L}^{\rm mean}}
    {\partial\kappa}\right|_0
    &=s(\boldy)-\langle s\rangle_S.
\end{align}
The Fisher information of the full likelihood for $N$ independent galaxies is therefore
\begin{equation}
    \calI_{\rm full}^{\rm mean}
    =N\Var_S(s)
    =\calI_{\rm opt}^{\rm mean}.
\end{equation}
Thus, the optimal sample mean captures all the Fisher information available from the galaxy-property distribution at fixed $N$.

\subsection{When and why the GFP residual is optimal}
The standard FP predicts the intrinsic size of galaxies from their surface brightness and central velocity dispersion $\sigma_{\rm v}$ through a linear relation,
\begin{equation}
    r_{\rm FP}=A_{\rm FP}\log_{10}\sigma_{\rm v}+B_{\rm FP}i+C_{\rm FP},
\end{equation}
We generalize this relation to $F(i,\bm{X})$, which is not restricted to a linear form and can incorporate additional lensing-invariant galaxy properties to minimize the intrinsic scatter. 

We now apply the general optimality results to the GFP residual. We first transform
the coordinates from $(r,m,\bm{X})$ to $(r,i,\bm{X})$ using
equation~(\ref{eq:SB}), and define the GFP coordinates as
$\bm{x}_{\rm FP}\equiv(i,\bm{X})$, where $\bm{x}_{\rm FP}$ is made of lensing-invariant quantities. The intrinsic GFP residual is
\begin{equation}
    \lambda_0(r,\bm{x}_{\rm FP})
    =r-F_0(\bm{x}_{\rm FP}),
    \label{eq:FP_residual}
\end{equation}
where $F_0$ denotes the intrinsic GFP relation before sample selection.
Since surface brightness and $\bm{X}$ are invariant under magnification,
lensing changes only $r$ at fixed $\bm{x}_{\rm FP}$. Specifically,
\begin{equation}
    5\left.\frac{\partial}{\partial m}\right|_r
    -\left.\frac{\partial}{\partial r}\right|_m
    =-\left.\frac{\partial}{\partial r}\right|_i,
\end{equation}
where $\bm{X}$ is held fixed throughout. The density score therefore becomes
\begin{equation}
    s(r,\bm{x}_{\rm FP})
    =-a\left.
    \frac{\partial\ln n_0(r,\bm{x}_{\rm FP})}{\partial r}
    \right|_{\bm{x}_{\rm FP}}.
\end{equation}

We can factorize the intrinsic number density as
\begin{equation}
    n_0(r,\bm{x}_{\rm FP})
    =n_0(\bm{x}_{\rm FP})\,
     p_0(r\mid\bm{x}_{\rm FP}),
\end{equation}
where $n_0(\bm{x}_{\rm FP})$ is the marginal number density at each
GFP position and $p_0(r\mid\bm{x}_{\rm FP})$ is the conditional
size distribution. Since $\bm{x}_{\rm FP}$ is invariant under
magnification, the score depends only on the conditional distribution,
\begin{equation}
    s(r,\bm{x}_{\rm FP})
    =-a\frac{\partial\ln p_0(r\mid\bm{x}_{\rm FP})}
             {\partial r}.
    \label{eq:conditional_score}
\end{equation}

If the intrinsic size distribution at fixed GFP coordinates is
Gaussian,
\begin{equation}
    p_0(r\mid\bm{x}_{\rm FP})
    =\frac{1}{\sqrt{2\pi}\sigma_0(\bm{x}_{\rm FP})}
    \exp\left[
    -\frac{[r-F_0(\bm{x}_{\rm FP})]^2}
           {2\sigma_0^2(\bm{x}_{\rm FP})}
    \right],
    \label{eq:gaussian_fp}
\end{equation}
equations~(\ref{eq:FP_residual}) and (\ref{eq:conditional_score})
give
\begin{equation}
    s(r,\bm{x}_{\rm FP})
    =\frac{a\lambda_0(r,\bm{x}_{\rm FP})}
           {\sigma_0^2(\bm{x}_{\rm FP})}.
    \label{eq:score_fp}
\end{equation}
According to equation~(\ref{eq:uniqueness}), the unique optimal
mean statistic, up to an affine transformation, is therefore
the inverse-variance weighted intrinsic GFP residual,
\begin{equation}
    f_{\rm opt}^{\rm mean}
    \propto\frac{\lambda_0}{\sigma_0^2(\bm{x}_{\rm FP})}.
    \label{eq:optimal_fp}
\end{equation}
In particular, if the intrinsic GFP scatter is constant, the
unweighted intrinsic GFP residual $\lambda_0$ itself is the unique optimal
mean statistic.

This result is independent of the marginal number density
$n_0(\bm{x}_{\rm FP})$, the functional form of $F_0$, and the
selection function $S$. The key requirement is that the intrinsic
conditional size distribution is Gaussian. For a non-Gaussian
distribution, the optimal observable can still be obtained
from equation~(\ref{eq:conditional_score}) using $\lambda_0$ distributions.

The response and Fisher information are also straightforward to
calculate. Defining
$g\equiv\lambda_0/\sigma_0^2(\bm{x}_{\rm FP})$, 
we obtain
\begin{align}
    \calR_g
    &=a\Var_S(g),\\
    \calI_g^{\rm mean}
    &=Na^2\Var_S(g).
    \label{eq:fp_general_info}
\end{align}
For constant intrinsic scatter, using $\lambda_0$ instead of $\lambda_0/\sigma_0^2$ as the estimator, we obtain
\begin{align}
    \calR_{\lambda_0}
    &=\frac{a}{\sigma_0^2}\Var_S(\lambda_0),
    \label{eq:fp_response}\\
    \calI_{\lambda_0}^{\rm mean}
    &=\frac{Na^2}{\sigma_0^4}\Var_S(\lambda_0).
    \label{eq:fp_optimal_info}
\end{align}
Without selection, $\Var(\lambda_0)=\sigma_0^2$ and hence
$\calR_{\lambda_0}=a$, recovering the pure size response.
Sample selection can modify the residual variance and therefore
both the magnification response and the information content,
even though the intrinsic GFP residual remains optimal. 

This result has a simple geometric interpretation. In the $(r,m)$
plane, magnification shifts galaxies along lines of constant surface
brightness $i$, while the perpendicular direction is invariant under
lensing. The magnification score is therefore determined by the
variation of the intrinsic galaxy distribution along the constant-$i$
direction. Conditioning on $i$ and other lensing-invariant properties
$\bm{X}$ isolates this variation, and a Gaussian conditional size
distribution makes the score proportional to the inverse-variance
weighted GFP residual. The additional properties $\bm{X}$ can help reduce
the intrinsic scatter $\sigma_0$, thereby increasing the available
magnification information.\footnote{In this context, the residual from the Kormendy relation, $r-r_{\rm Kmndy}(i)$, generally contains less magnification information than that from the traditional FP, $r-r_{\rm FP}(i,\sigma_{\rm v})$, which additionally incorporates velocity dispersion. Including further lensing-invariant properties can improve the magnification precision even more.} In this sense, the GFP provides an optimal compression of the full galaxy-property likelihood by separating lensing-invariant and lensing-dependent directions.

Importantly, this also shows that GFP magnification combines information
from both flux and size, rather than size alone as is often assumed.
Although surface brightness itself is invariant under lensing, the
GFP uses both flux and size through their combination in $i$.
Indeed, we can equivalently use the GFP to predict magnitude instead
of size. Defining the magnitude residual at fixed $(i,\bm{X})$ gives
\begin{equation}
    m-m_{\rm FP}(i,\bm{X})=-5[r-F_0(i,\bm{X})]=-5\lambda_0,
\end{equation}
which contains exactly the same magnification information as the
size residual. 

\subsection{GFP fitting in a  selection-biased sample}

Although sample selection does not affect the optimality of the
intrinsic GFP residual $\lambda_0$,
in practice the GFP is often fitted directly to the selected sample,
yielding a different residual $\lambda_S$. This raises two questions:
whether $\lambda_S$ remains optimal and, if not, whether the optimal
estimator can be recovered using only the selected sample.

Equation~(\ref{eq:conditional_score}) provides a straightforward
answer. At fixed $\bm{x}_{\rm FP}$, the conditional distribution
of the selected sample is
\begin{equation}
    p_S(r\mid\bm{x}_{\rm FP})
    \propto S(r,\bm{x}_{\rm FP})\,
    p_0(r\mid\bm{x}_{\rm FP}).
\end{equation}
The intrinsic score can therefore be expressed as
\begin{equation}
    s=-a\frac{\partial\ln p_S(r\mid\bm{x}_{\rm FP})}{\partial r}
      +a\frac{\partial\ln S(r,\bm{x}_{\rm FP})}{\partial r}.
    \label{eq:selected_score}
\end{equation}
For a sharp selection, $S$ is constant within the selected
region, and the second term vanishes away from the selection
boundaries. The optimal estimator can then be constructed
directly from the conditional distribution of the selected
sample, without requiring knowledge of the intrinsic GFP
relation $F_0$. This result holds for any intrinsic
distribution, not necessarily Gaussian.

For a Gaussian intrinsic GFP residual, the optimal score can be estimated
more directly by exploiting the known functional form of the
conditional distribution. Defining the selected GFP residual
as $\lambda_S=r-F_S(\bm{x}_{\rm FP})$, the logarithm of its
conditional distribution under a sharp selection remains
quadratic within the selected region,
\begin{equation}
    \ln p_S(\lambda_S\mid\bm{x}_{\rm FP})
    =C(\bm{x}_{\rm FP})
    +b_1(\bm{x}_{\rm FP})\lambda_S
    -\frac{1}{2}b_2(\bm{x}_{\rm FP})\lambda_S^2,
\end{equation}
where $F_S$ is the GFP fitted to the selected sample. The optimal
score is therefore
\begin{equation}
    s=a\left[b_2(\bm{x}_{\rm FP})\lambda_S
    -b_1(\bm{x}_{\rm FP})\right],
    \label{eq:selected_gaussian_score}
\end{equation}
where $b_2=1/\sigma_0^2$ accounts for the intrinsic scatter,
while $b_1=(F_0-F_S)/\sigma_0^2$ corrects for the
selection-induced shift of the GFP. Both coefficients can be
determined directly by fitting a truncated Gaussian to the
selected sample, accounting for the selection boundaries.
Thus, $\lambda_S$ alone is generally not optimal, but the
optimal estimator remains a linear function of $\lambda_S$
at fixed GFP coordinates.

Compared with directly estimating the logarithmic derivative
in equation~(\ref{eq:selected_score}), the Gaussian approach
avoids numerical differentiation and can provide a more stable
estimate of the score for finite samples. However, it relies
on the Gaussian assumption, while the direct approach applies
to general distributions. The choice between them therefore
depends on the accuracy of the Gaussian approximation and the
available sample size.

\section{Summary Statistics}
\label{sec:summary}

In the previous section, we derived the optimal magnification estimator based on the sample mean of a galaxy observable. However, magnification also changes the total number of observed galaxies, providing additional information. We now extend the analysis to more general summary statistics that also account for changes in number counts.

For a population of independent galaxies following Poisson statistics, we consider the sum of an arbitrary observable $f(\bm{y})$ over all observed galaxies,
\begin{equation}
    T_f=\sum_{j=1}^{N}f(\bm{y}_j),
    \label{eq:summary_statistic}
\end{equation}
where $N$ is the observed number of galaxies, which is allowed to fluctuate. We denote the expected number of galaxies without magnification as
\begin{equation}
    \bar{N}_0^S=\Omega\int\dd\bm{y}\,S(\bm{y})n_0(\bm{y}),
\end{equation} 
where $\Omega$ is the survey area. Unlike the sample mean considered previously, the variance of $T_f$ includes both the intrinsic scatter of $f$ and fluctuations in the number of galaxies. The expectation and variance are therefore
\begin{align}
    \langle T_f\rangle_\kappa
    &=\Omega\int\dd\bm{y}\,f(\bm{y})n^S_\kappa(\bm{y}),\\
    \Var(T_f)
     &=\bar{N}_0^S\Var_S(f)+\bar{N}_0^S\langle f\rangle_S^2 
    =\bar{N}_0^S\langle f^2\rangle_S \nonumber\\
    &=\Omega\int\dd\bm{y}\,f^2(\bm{y})S(\bm{y})n_0(\bm{y}),
\end{align}
where the variance is evaluated at $\kappa=0$, and we have used the Poisson relation $\Var(N)=\bar{N}_0^S$.

Using equation~(\ref{eq:population}), the magnification
response of $T_f$ is
\begin{equation}
    \left.\frac{\partial\langle T_f\rangle_\kappa}
    {\partial\kappa}\right|_0
    =\bar{N}_0^S\,\langle f(s-2)\rangle_S.
\end{equation}
The corresponding local inverse variance is
\begin{equation}
    \calI_f^{\rm sum}
    =\bar{N}_0^S\frac{\langle f(s-2)\rangle_S^2}
                    {\langle f^2\rangle_S}.
    \label{eq:summary_fisher}
\end{equation}
Applying the Cauchy--Schwarz inequality gives
\begin{equation}
    \calI_f^{\rm sum}
    \leq \bar{N}_0^S\langle(s-2)^2\rangle_S,
    \label{eq:summary_bound}
\end{equation}
with equality if and only if
\begin{equation}
    f_{\rm opt}^{\rm sum}(\bm{y})
    \propto s(\bm{y})-2.
    \label{eq:optimal_summary}
\end{equation}
Unlike the optimal mean statistic, the optimal sum
statistic is unique up to a multiplicative constant,
as an additive constant contributes to the number counts.

This result is again equivalent to a full-likelihood analysis. For a Poisson galaxy population, the likelihood of the observed galaxy distribution is
\begin{align}
    \mathcal{L}^{\rm sum}(\kappa\mid N,\{\bm{y}_j\})
    &=P(N|\kappa)\prod_{j=1}^{N}p_\kappa^S(\bm{y}_j) 
    \propto e^{-\bar N_\kappa^S}
    \prod_{j=1}^{N}n_\kappa^S(\bm{y}_j),
    \label{eq:summary_likelihood}
\end{align}
where $\bar{N}_\kappa^S=\Omega\int\dd\bm{y}\,n_\kappa^S(\bm{y})$ is the expected number of selected galaxies under magnification. The corresponding likelihood score is
\begin{equation}
    \left.\frac{\partial\ln\mathcal{L}^{\rm sum}}
    {\partial\kappa}\right|_0
    =\sum_{j=1}^{N}[s(\bm{y}_j)-2]
    -\bar{N}_0^S\langle s-2\rangle_S.
    \label{eq:summary_likelihood_score}
\end{equation}
For $f_{\rm opt}^{\rm sum}=s-2$, this is simply $T_{\rm opt}-\langle T_{\rm opt}\rangle_0$. Therefore,
\begin{equation}
    \calI_{\rm full}^{\rm sum}
    =\bar{N}_0^S\langle(s-2)^2\rangle_S=\calI_{\rm opt}^{\rm sum}.
    \label{eq:summary_full_information}
\end{equation}

To separate the contributions from the mean statistic and number counts, we define the
magnification response of the total number counts as
\begin{equation}
    \calR_N
    \equiv\left.\frac{\partial\ln \bar{N}_\kappa^S}
    {\partial\kappa}\right|_0
    =\langle s\rangle_S-2.
    \label{eq:number_count_response}
\end{equation}
The optimal observable can then be decomposed as
\begin{equation}
    f_{\rm opt}^{\rm sum}(\bm{y})
    =\underbrace{s(\bm{y})-\langle s\rangle_S}_
    {f_{\rm opt}^{\rm mean}(\bm{y})}
    +\,\calR_N,
\end{equation}
where we have chosen the mean-subtracted form of
$f_{\rm opt}^{\rm mean}$. 
The corresponding information is
\begin{equation}
    \calI_{\rm opt}^{\rm sum}
    =\calI_{\rm opt}^{\rm mean}
    + \bar{N}_0^S\,\calR_N^2,
    \label{eq:information_decomposition}
\end{equation}
where $\calI_{\rm opt}^{\rm mean}=\bar{N}_0^S\Var_S(s)$
is averaged over the Poisson-distributed sample size.
Therefore, the optimal summary statistic is equivalent
to combining the optimal mean statistic with total number
counts, which provide additional information whenever
$\calR_N\neq0$.

As shown in Section~\ref{sec:mean}, for a Gaussian intrinsic
GFP residual, the optimal mean observable is the inverse-variance-weighted GFP residual, reducing to the ordinary GFP residual for constant intrinsic scatter. Thus, under the same assumptions, the full magnification information from galaxy surveys can be captured by the GFP residual and total number counts.

It is also worth noting that previous studies have optimized number-count magnification measurements by weighting galaxies according to their magnitude-dependent responses \citep{2002A&A...386..784M,2005PhRvL..95x1302Z,2016MNRAS.456.2518G}. Our results show that the additional information gained from such weighting is part of the galaxy-property information already captured by the optimal mean statistic. Therefore, once the optimal mean statistic is combined with the total number counts, further weighting of the counts provides no additional information.

\section{Discussion}\label{sec:con}

In this Letter, we have shown that the optimal summary statistic for weak-lensing magnification captures all the Fisher information available from galaxy properties $(r,m,\bm{X})$ and number counts, achieving the same precision as a full-likelihood analysis. This statistic can be decomposed into the optimal mean statistic and the total number counts. We demonstrate that the optimal mean statistic can be constructed from the conditional GFP residual distribution. For a Gaussian intrinsic GFP residual, it reduces to the inverse-variance-weighted GFP residual, or simply the intrinsic GFP residual when the scatter is constant.

Our results establish the GFP residual and number counts as an optimal compression of the full galaxy-property likelihood. This is particularly useful because incorporating additional lensing invariant galaxy properties $\bm{X}$ can reduce the GFP scatter and increase the available magnification information, while making the full likelihood increasingly high-dimensional and difficult to model. The GFP provides a much simpler representation without losing information. Moreover, the GFP offers an intuitive physical interpretation of magnification measurements, combining information from both galaxy flux and size. As the FP has been extensively studied in other contexts, existing knowledge of galaxy scaling relations can also help improve magnification measurements.

Our analysis assumes an idealized galaxy population with independent Poisson statistics, neglecting measurement errors in $(r,m,\bm{X})$ and intrinsic galaxy clustering. These effects can modify the optimal weighting and information content, potentially depending on the specific measurement, such as magnification autocorrelations or galaxy--galaxy lensing \citep{2005PhRvL..95x1302Z}. We also leave the choice of $\bm{X}$ unspecified, as the optimal set of lensing-invariant properties depends on the available photometric and spectroscopic information. Although our derivation is formulated directly in terms of observed galaxy properties and does not require explicit redshift information, the GFP relation may depend on redshift through galaxy evolution, k-corrections, and cosmological surface-brightness dimming. Photometric-redshift uncertainties can therefore mix populations with different GFP relations, potentially altering the residual distribution and the optimal weights. Such uncertainties also affect the lensing geometry when interpreting magnification measurements in terms of the underlying matter distribution.
Nevertheless, our results provide a useful theoretical benchmark for understanding and optimizing weak-lensing magnification measurements.

\section*{Acknowledgements}
K.X. is supported by funding from the Center for Particle Cosmology at U Penn.

\section*{Data Availability}
No new data were generated or analysed in support of this research.



\bibliographystyle{mnras}
\bibliography{main} 







\bsp	
\label{lastpage}
\end{document}